\documentclass[prl,twocolumn,showpacs,showkeys,preprintnumbers,floatfix,superscriptaddress,longbibliography]{revtex4-1}

\usepackage{amsfonts} 
\usepackage{amssymb} 
\usepackage{amsmath} 
\usepackage{graphicx} 
\usepackage{subfigure} 
\usepackage{array} 
\usepackage{dcolumn} 
\usepackage{bm} 
\usepackage{latexsym} 
\usepackage{longtable} 
\usepackage{hyperref} 
\usepackage{bbold}
\usepackage{mathtools}
\usepackage{soul}

\usepackage[utf8]{inputenc}
\usepackage[braket]{qcircuit}
\usepackage[dvipsnames,usenames]{xcolor}
\usepackage[normalem]{ulem}

\newcommand{\Bv}{\vec{B}}
\newcommand{\vvec}{\mathbf{v}}

\newcommand{\sigv}{\vec{\sigma}}

\newcommand{\qmouter}[2]{ \rvert #1 \rangle \! \langle #2 \rvert}

\newcommand{\UNITN}{{Dipartimento di Fisica, University of Trento, via Sommarive 14, I–38123, Povo, Trento, Italy}}
\newcommand{\TIFPA}{INFN-TIFPA Trento Institute of Fundamental Physics and Applications,  Trento, Italy}
\newcommand{\LANL}{Theoretical Division, Los Alamos National Laboratory, Los Alamos, New Mexico, 87545}
\newcommand{\UNM}{Department of Physics and Astronomy, University of New Mexico, Albuquerque, New Mexico 87131, USA}

\begin{document}
\title{Many-body neutrino flavor entanglement in a simple dynamic model}
\author{Joshua D. Martin}
\affiliation{\LANL}
\author{A. Roggero}
\affiliation{\UNITN}
\affiliation{\TIFPA}
\author{Huaiyu Duan}
\affiliation{\UNM}
\author{J. Carlson}
\affiliation{\LANL}

\preprint{LA-UR-23-20394}

\date{\today}

\begin{abstract}
    Dense neutrino gases form in extreme astrophysical sites, and the flavor content of the neutrinos likely has an important 
    impact on the subsequent dynamical evolution of their environment. Through coherent forward scattering among neutrinos, the flavor content of the 
    gas evolves under a time-dependent potential which can be modeled in a quantum many-body formalism as an all-to-all coupled spin-spin interaction. 
    This two-body potential generically introduces entanglement and greatly complicates the study of these systems.  In this work we study the evolution 
    of the quantum many-body problem as well as the typically employed mean-field approximation to it for a small number of neutrinos ($N = 16$).  We 
    consider randomly chosen one- and two-body couplings in the Hamiltonian, and the resulting evolution of several
    initial product states. We subsequently compare many-body and mean-field predictions for one-body observables, and we consider one- and two-body entanglement 
    to assess under what conditions the many-body and mean-field predictions are likely to disagree. Except for a special category of 
    prototypical initial conditions, we find that the typically employed mean-field approximation is insufficient to capture the evolution of one-body 
    operators in the systems we consider. We also observe a loss of coherence in one- and two-body trace-reduced subsystems which suggests that the evolution may be well approximated as a classical mixture of separable states.  
\end{abstract}

\maketitle

{\it Introduction--- } In hot and dense astrophysical environments such as core collapse supernovae (CCSN) and binary neutron star mergers (BNSMs), neutrinos are emitted in such great numbers that their dynamical evolution is important to the overall evolution of their local environment.  In CCSN explosions neutrinos are expected to be an important component of the chemical and hydrodynamic evolution of the explosion, while for both CCSN and BNSMs the neutrino gas impacts the local proton-to-neutron ratio thereby affecting the r-process nucleosynthesis of heavy elements \cite{Bethe_RevModPhys.62.801,pantaleone1992neutrino,Janka:2006fh,Woosley:2005,Hoffman:1997,li2021neutrino,Fernandez:2022yyv}.  As neutrinos change flavor due to vacuum oscillations and through charged and neutral current coherent forward scattering with particles in a dense medium, accurately modeling their flavor evolution is critical to obtaining a detailed understanding of these extreme astrophysical events \cite{Sigl:1992fn,Qian:1994wh,Qian:1995,pastor2002flavor,pastor2002physics,Bell:2003mg,sawyer2004classical,Balantekin:2006tg}.

Of particular importance is the effect of flavor exchange through neutral current coherent forward scattering among local neutrinos on different trajectories \cite{Sigl:1992fn,pantaleone1992neutrino,Qian:1994wh,Qian:1995,pastor2002flavor,pastor2002physics}.  This process results in a coupling of the neutrino gas to itself by correlating the evolution histories of neutrinos along different trajectories.  These correlations in turn makes tracking the detailed flavor evolution of the gas challenging even in simplified models.  A direct quantum many-body treatment requires evolving a number of amplitudes which scales exponentially in the number of neutrinos, thus approximations are typically employed.  One common approximation is a mean-field (MF) treatment of the $\nu-\nu$ coherent forward scattering potential under which only expectation values of one-body operators are considered.  However, recent work has been unclear regarding whether such a treatment produces consistent predictions for one-body expectation values as compared with a formalism which retains all quantum many-body (MB) flavor correlations \cite{Birol:2018qhx,Patwardhan2019,Rrapaj2020,Roggero2021a,Roggero2021b,Xiong:2021dex,Martin:2021bri,Roggero2022,Siwach:2022xhx,Lacroix:2022krq}.

One predicted feature ubiquitous among MF treatments of the flavor evolution is that of spectral splits and swaps \cite{Raffelt:2007cb,Raffelt:2007xt}. This is when neutrinos beginning in one flavor state in some window of energy are adiabatically converted to a different flavor state resulting in a swap of the flavor spectra between neutrino species with an associated sharp feature in the spectra at the energy window boundaries.  Such behavior has also been predicted and observed \cite{Birol:2018qhx,patwardhan2021spectral} in the MB treatment within the single-angle approximation.  This approximation has the advantage that it results in an integrable Hamiltonian and exact solutions are possible via a Bethe Ansatz.

In this letter, we will study the MB evolution of a small system of 16 interacting neutrinos and compare the resulting behavior under the MF approximation. We will adopt the two flavor approximation thus representing the flavor state of each neutrino as an SU(2) spinor, though recent work has demonstrated the importance of considering three flavor evolution \cite{Siwach:2022xhx}.

The Hamiltonian we will study has the form~\cite{Pehlivan2011}
\begin{equation}\label{eq:totalH} 
    H = H_{\text{vac}} + H_{\nu \nu}(t)
\end{equation}
where
\begin{equation} \label{eq:Hvac}
    H_{\text{vac}} = \sum_{i=1}^{N} \frac{\omega_{i}}{2} \Bv \cdot \sigv_{i} \, .
\end{equation}
In the mass basis for normal mass ordering, $\Bv\!\!=\!\! (0,0,-1)$.  In choosing the vacuum oscillation frequencies, 
we construct $N=16$ bins of width $\omega_{0}$, where $\omega_{0}$ is a reference vacuum oscillation frequency which controls 
the scale of the one-body Hamiltonian.  For a 10 MeV neutrino, $\omega_{0} \approx 1$ km$^{-1}$. Then for the one-body 
couplings for each neutrino, we choose $\omega_{i}$ randomly from the uniform interval
$\omega_{0} \left[ i-1, i \right]$.

The neutrino-neutrino interaction Hamiltonian is SU(2) invariant and takes the form of an all-to-all coupled Heisenberg model
\begin{equation} \label{eq:manyBodyH}
    H_{\nu \nu}(t) = \frac{\mu(t)}{2 N} \sum_{i < j} 
        \left(1 - \vvec_{i} \cdot \vvec_{j} \right) \sigv_{i} \cdot \sigv_{j} \,  .
\end{equation}
Here we employ the parameterization of $\mu(t)$ from~\cite{patwardhan2021spectral}
\begin{equation}
    \mu(t) = \mu_{0} \left(1 - \sqrt{1 - \left( \frac{R_{\nu}}{r_{0} + t} \right)^2 } \right)^{2}
\end{equation}
with $R_{\nu} = 32.2 \omega_{0}^{-1}$, and $r_{0} = 210.64 \omega_{0}^{-1}$.  This choice of $\mu(t)$ was derived under the 
single angle approximation, but the geometric corrections from multi-angle effects contribute to the structure of the potential at higher orders in an expansion of $\mu(t)$ in powers of $R_{\nu} / r$, so we have 
retained its form for ease of comparison.  We choose $\mu_{0}$ such that
$\mu(t=0) = 80 \omega_{0}$.  The two-body Hamiltonian we employ differs by a factor of $1/N$ relative to that of \cite{patwardhan2021spectral}, 
thus we choose $\mu(t=0)$ a factor of 16 larger than \cite{patwardhan2021spectral} in our definition of $\mu(t)$ in order to match the Hamiltonian employed by that work which employs uniform couplings (UC) with
$\vvec_{i} = 0$. We also consider the case of random couplings (RC) by choosing the neutrino velocities in Eq.~\eqref{eq:manyBodyH} as
$\vvec_{i} = \left[ \cos(\theta_{i}),0,\sin(\theta_{i}) \right]$, with
$\theta_{i}$ chosen randomly from the uniform interval
$\frac{\pi}{N}\left[ i-1 , i \right]$.
The choice of selecting frequencies $\omega_i$ and angles $\theta_i$ with a random component ensures no accidental energy degeneracy is produced as a result of the necessary, but arbitrary, discretization procedure.

We will study initial conditions of the form
\begin{equation} \label{eq:IC_state}
    \ket{\Psi(t=0;n)} = \bigotimes_{i=1}^{n} \ket{\nu_{e}}_{i} \bigotimes_{j=n+1}^{N} \ket{\nu_{\tau}}_{j} \;,
\end{equation}
where
$\ket{\nu_e}=\cos\theta_\text{vac} \ket{\nu_1} + \sin\theta_\text{vac}\ket{\nu_2}$, and
$\ket{\nu_\tau}= - \sin\theta_\text{vac} \ket{\nu_1} + \cos\theta_\text{vac}\ket{\nu_2}$.
Here $\ket{\nu_1}$ and $\ket{\nu_2}$ are the mass eigenstates of the neutrino, and $\theta_{\text{vac}} = 0.584$ is the vacuum mixing angle.

The Hamiltonian specified in Eq.~\eqref{eq:totalH} commutes with the total projected angular momentum in the mass basis $J_{z} = \frac{1}{2} \sum_{i=1}^{N} \sigma_{z}^{(i)}$
and thus the projections of the quantum state into the different $J_{z}$ invariant subspaces evolve fully decoupled from one another.
Also, for any choice of couplings, the total squared angular momentum operator $J^{2} = \frac{1}{4} \sum_{i,j=1}^{N} \sigv_{i} \cdot \sigv_{j}$
commutes with the two-body Hamiltonian $H_{\nu \nu}$ due to the SU(2) invariance of the latter and therefore the two-body interaction cannot connect states of different total angular momentum.

{\it Fully polarized initial product states---}
Any initial condition which is populated by product states of all equal polarization is an eigenstate of $J^{2}$ with maximum $J$ quantum number.  
Because $J^{2}$ commutes with $H_{\nu \nu}$, this state is also approximately an eigenstate of $H$ at early time when $H_{\nu \nu} \gg H_{\text{vac}}$.
Once evolution begins, and assuming that in each invariant subspace the evolution remains adiabatic, then the final flavor configuration 
(at $\mu=0$) will be of the form of a split-state
which is a linear combination of the highest energy state of $H_{\text{vac}}$ in each $J_z$ invariant subspace~\cite{Birol:2018qhx}. 

\begin{figure}
    \centering
    \includegraphics[scale=0.31]{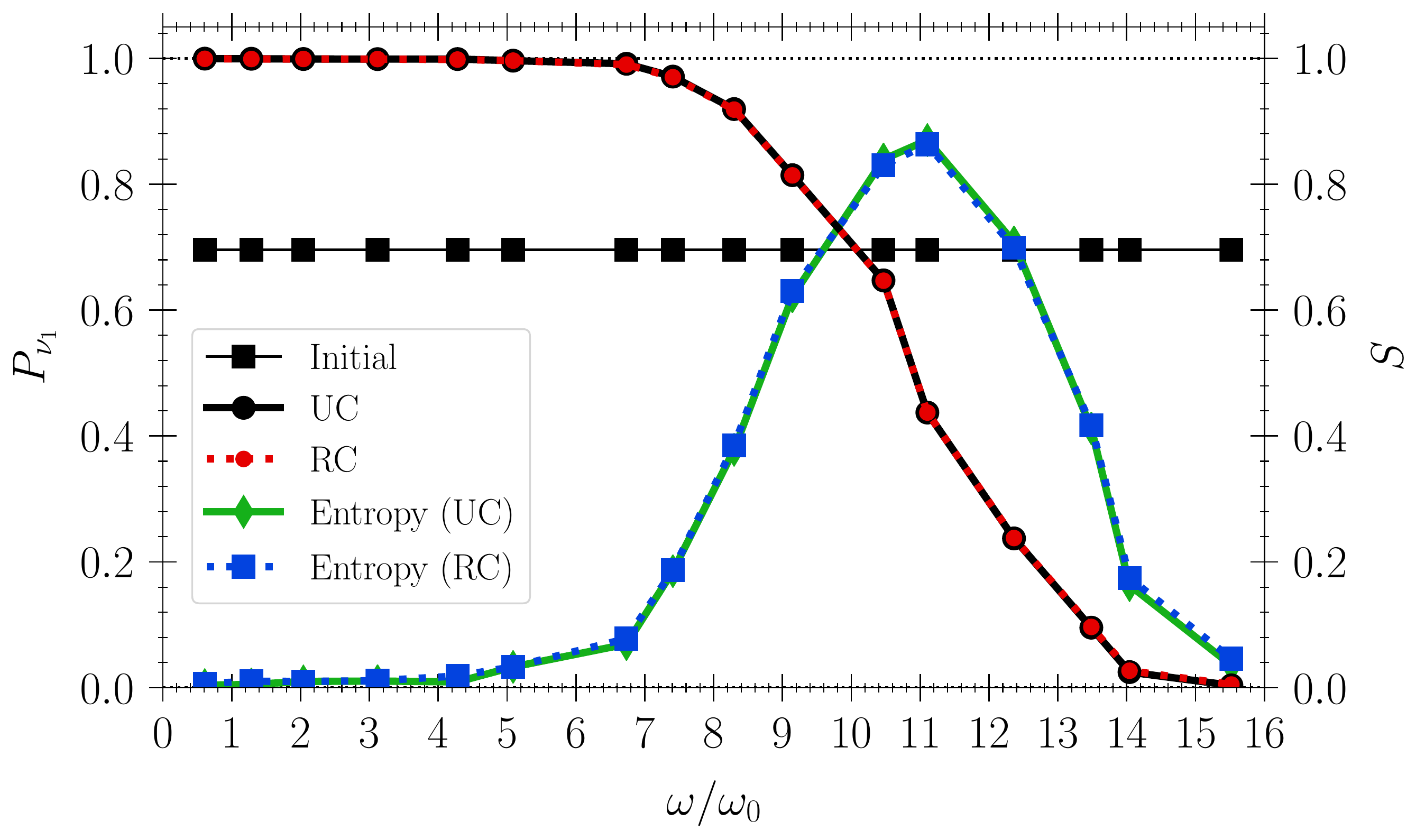}
    \caption{
    The initial and final probabilities $P_{\nu_1}$ of each neutrino in $\ket{\nu_1}$ for the fully polarized initial state 
    and with the uniform angular coupling (UC) and random couplings (RC), respectively. Also plotted are the final one-body 
    entropy $S$ in both scenarios.
    }
    \label{fig:allNuE_vBinning_comparison}
\end{figure}

The possibility of adiabatic evolution for this state 
 with UC has been observed in past  work employing Bethe-Ansatz techniques~\cite{Birol:2018qhx,Cervia2019}. In order to validate this analysis for general angular configurations, 
we choose a state fully polarized as $\nu_{e}$ initially
[$n=N$ in Eq.~\eqref{eq:IC_state}], and we compute the expectation values of $\sigma_{z}$ in the mass basis 
as well as the entanglement entropy, $S(\rho_i)=-\textrm{Tr}[\rho_i\log_2(\rho_i)]$, of the one-body reduced density matrices (RDMs) with both UC and RC. The results of 
these two choices of the two-body couplings are shown in Fig.~\ref{fig:allNuE_vBinning_comparison}. 

The results in Fig.~\ref{fig:allNuE_vBinning_comparison} are
fully compatible with the conclusions of Ref.~\cite{patwardhan2021spectral}, where adiabaticity can be 
justified thanks to the integrability of the model with uniform couplings. We also have not observed a violation of the 
adiabaticity of the evolution for uniformly polarized initial states for any choice of the velocities.  However, should adiabaticity 
violation occur for some special choice of one and two-body couplings, the final state should not be expected to take the asymptotic split-state
form as above.

{\it Partially polarized product states---}
When we consider an initial condition with some number of $\ket{\nu_{\tau}}$ states in Eq.~\eqref{eq:IC_state}, we no longer begin 
in the highest energy states of each $J_{z}$ subspace.  Initially the highest energy state in the individual $J_{z}$ subspaces 
are states with $\ket{J^{2} = \frac{N}{2} \left(\frac{N}{2} + 1 \right);J_z=m}$.  To form a product state with $n \neq N$  neutrinos in 
the $\nu_{e}$ state, we require linear combinations of states from all $J^{2}$ subspaces with $J \geq |n - \frac{N}{2}|$.  Thus the 
initial condition will be distributed in a given $J_{z}$ subspace across a range of initial energy states. To investigate the dynamics 
which arise for this category of initial condition we choose all neutrinos with $\omega_{i} > 12$ as $\ket{\nu_{\tau}}$ 
flavor states.

\begin{figure} 
    \centering 
    \includegraphics[scale=0.33]{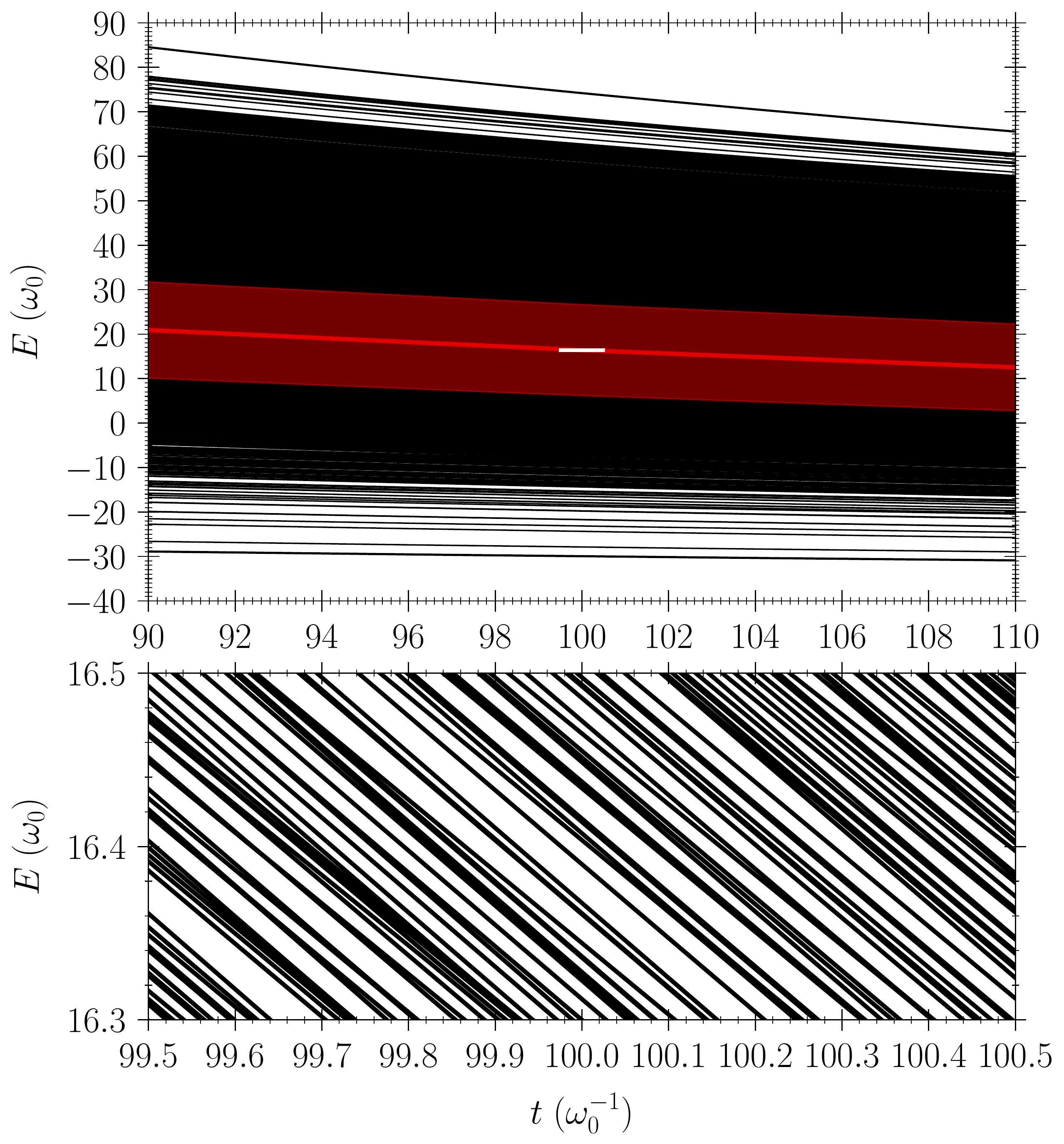} 
    \caption{Top panel: evolution of the full energy spectrum in a narrow window of time for the 
    $m=1$ subspace (with 9 \ket{\nu_{1}} states). Bottom panel: The evolution of the spectrum zoomed 
    into the energy and time range indicated by the white rectangle in the top panel.} 
    \label{fig:energy_spec} 
\end{figure} 

The time dependent Hamiltonian generically exhibits avoided level crossings in its energy spectrum at times for which $H_{\text{vac}} \sim H_{\nu \nu}$. Transiting these crossings 
nonadiabatically will result in evolution of the state in the spectrum, and there is no \textit{a priori} reason to believe that under 
such conditions an initial product state will remain approximately a product state.
We can explicitly consider the spectrum of the RC Hamiltonian in the $m=1$ $J_z$ subspace in Fig.~\ref{fig:energy_spec}. The black lines 
show the evolution of the energy eigenvalues as a function of time in this subspace. We note that the gap between the highest 
energy state and the rest of the spectrum is consistent with the observation that fully polarized initial conditions (as in the previous section) evolve 
adiabatically in the highest energy state in each invariant subspace. In the top panel the red line represents the average energy in this subspace of 
the example state under consideration, while the red band about the mean represents the one standard 
deviation width of the state in the energy spectrum of this $J_z$ invariant subspace.

The white bar centered at $t=100 \omega_{0}^{-1}$ and $E = 16.4 \omega_{0} $ in the top panel
of Fig.~\ref{fig:energy_spec}
is a window in time and energy in the spectrum 
which we show in detail in the bottom panel.  This shows the presence of a large number of avoided level 
crossings in the time evolution of the spectrum.  In the absence of an extensive set of conserved charges for the RC case, we do not 
expect that these crossings can be transited adiabatically as was argued for the UC case in Appendix B of \cite{Cervia2019}.

\begin{figure} 
    \centering 
    \includegraphics[scale=0.35]{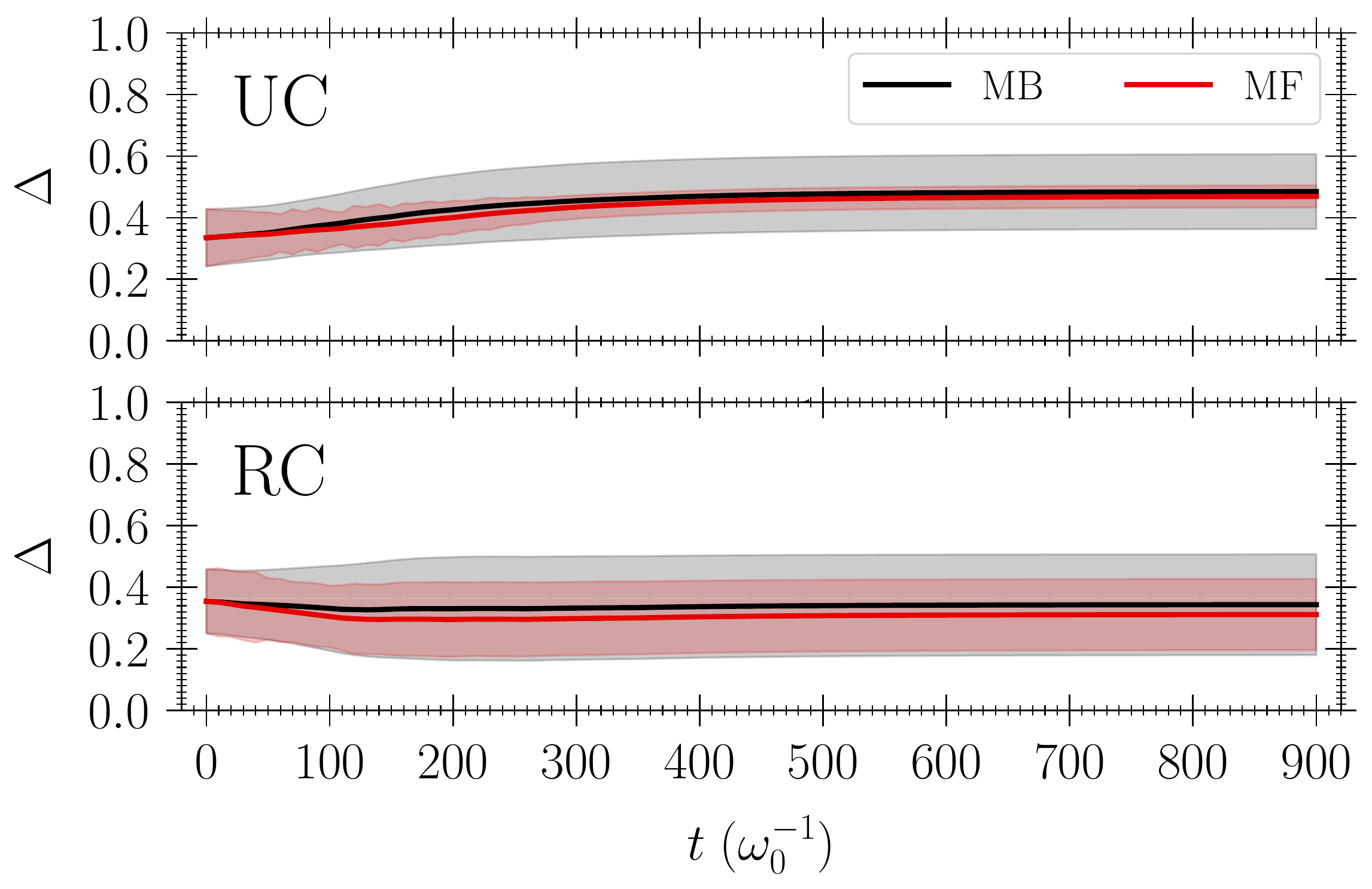} 
    \caption{Evolution of the MB and MF neutrino distributions relative to the total Hamiltonian spectral width.  
    The black (red) lines represent the average energy of the MB (MF) system, while the grey (red) band represents 
    the one-sigma width of the distribution in the spectrum.} 
    \label{fig:topQuarterTau_deltaSpect} 
\end{figure} 

During the evolution, a necessary but not sufficient condition for good agreement between one-body operator expectation values of the 
MF and MB formalisms is that both the average energy and the variance of the vacuum Hamiltonian be in agreement. 
We observe, however, 
that differences in both the expectation value and the variance of the total Hamiltonian accumulate between the MB and MF solutions 
during the time evolution of the system which we show in Fig.~\ref{fig:topQuarterTau_deltaSpect}. At each time, we diagonalize the 
Hamiltonian and recover the largest and smallest energy eigenvalues $E_{\textrm{max}}$ and $E_{\textrm{min}}$.  We then calculate the difference between the quantum state's average
energy and $E_{\textrm{min}}$ relative to the total width of the spectrum, 
i.e. 
\begin{equation}
    \Delta = \frac{ \langle H \rangle - E_{\textrm{min}} }{ E_{\textrm{max}} - E_{\textrm{min}} } \, .
\end{equation}
We similarly compute the standard deviation of $H$ and normalize it to the total width of the spectrum at each time in order to get a 
sense of how many energy states potentially have overlap with the system.  If, during the evolution, the MB system transits avoided level 
crossings in the spectrum nonadiabatically, it is a generic expectation that the variance of the Hamiltonian relative to the width of the 
spectrum should increase in time for the interval over which the spectrum is appreciably dynamic.
This is indeed what is observed in Fig.~\ref{fig:topQuarterTau_deltaSpect}.

For both models at late time, the MB solution average energy has separated from the MF solution which implies a difference in the polarizations of the neutrinos in both the mass and flavor bases. The width of the state in the energy spectrum of the vacuum Hamiltonian also indicates that the mass state 
distributions must vary between the MB and MF cases.  
This is because,
in the $\mu \rightarrow 0$ limit  relevant to late times, the variance of the Hamiltonian is
\begin{align}
    \Delta H_{\text{vac}}^{2} =& \frac{1}{4} \sum_{i} \omega_{i}^{2} \left(1 - \langle \sigma_{z}^{(i)} \rangle^{2} \right) \nonumber \\
        &+ \frac{1}{2} \sum_{i<j} \omega_{i}\omega_{j} 
            \left( \langle \sigma_{z}^{(i)} \sigma_{z}^{(j)} \rangle - \langle \sigma_{z}^{(i)} \rangle \langle \sigma_{z}^{(j)} \rangle \right) \, .
\end{align}
Thus the variance is sensitive to both one-body expectation values and two-body quantum correlations.
For the MF case, the two body term is zero by assumption, and it also vanishes for a statistically mixed state.
We note, however, that the degree of disagreement for $\langle \sigma_{z}^{(i)} \rangle$ is 
not directly inferable from the first and second moments of the vacuum Hamiltonian at late time.

\begin{figure}
    \centering
    \includegraphics[scale=0.33]{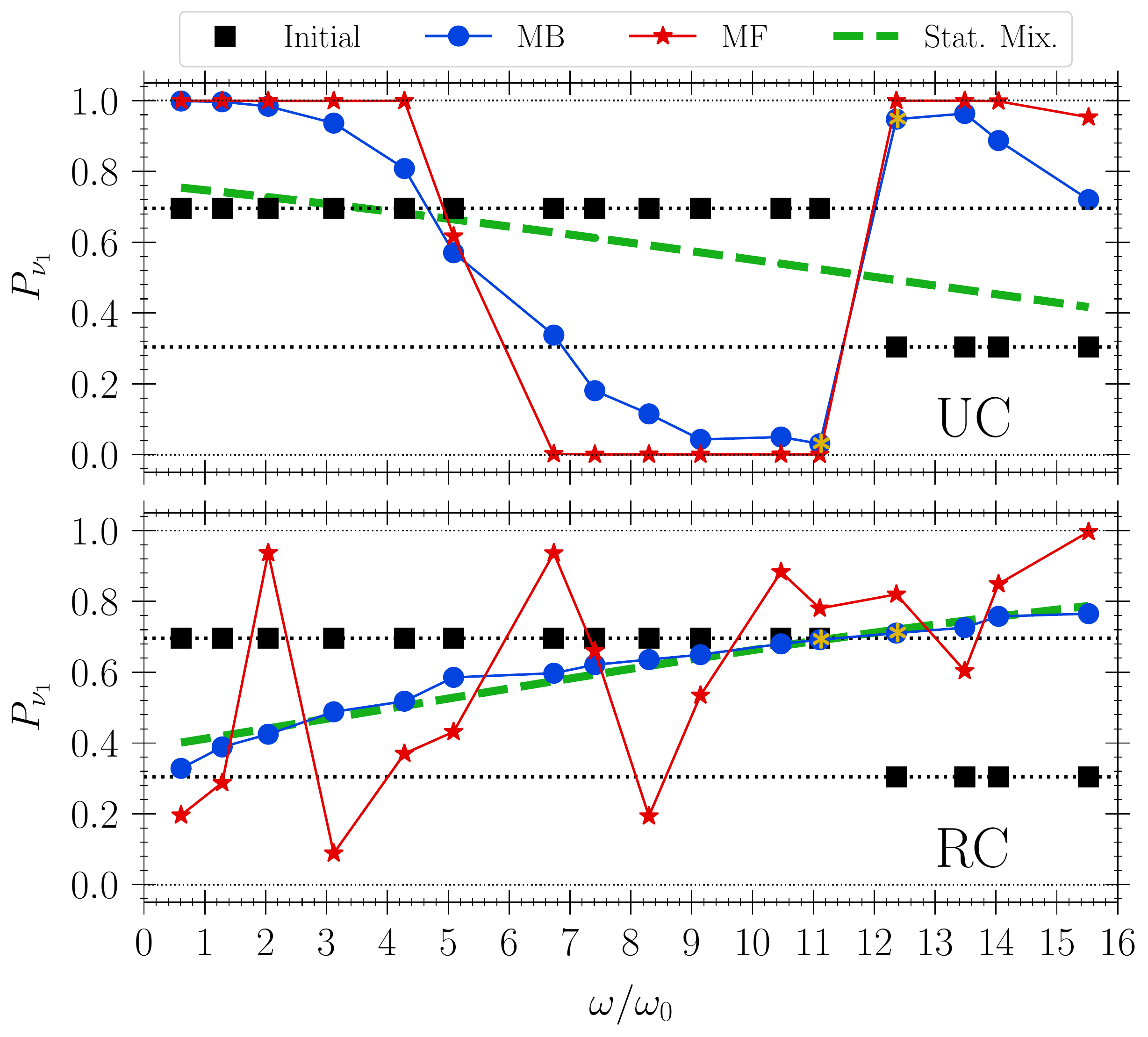}
    \caption{  The initial and final probabilities $P_{\nu_1}$ of each neutrino in the 
    MF approximation and the full 
    MB calculation with the uniform coupling (upper panel) and random couplings (lower panel), respectively. The dashed curves are for the statistical mixture states described by Eq.~\eqref{eq:statFit_1}. 
    The two gold asterisks indicate the neutrinos which are included in the one- and two-body trace distance plots in Fig.~\ref{fig:trDistances}.
    }
    \label{fig:obFlavorOps}
\end{figure}

In Fig.~\ref{fig:obFlavorOps} we show the late-time $\nu_{1}$ measurement probability in both the MF and MB formalisms with the two different choices of couplings. 
For the UC case (top panel), the final state shows qualitatively good 
agreement between the MB and MF formalisms.  Furthermore, we observe and confirm (without plotting) that the one-body 
von Neumann entropy is largest at the split location which was a key result of~\cite{patwardhan2021spectral}.

In contrast, in the RC case we observe substantial disagreement between the MF and MB predictions for the $\nu_{1}$ polarization 
(bottom panel of Fig.~\ref{fig:obFlavorOps}). We also observe a near total loss of coherence between the mass states in the MB 
formalism. This is because, 
when the initial condition is extended 
in the spectrum of the Hamiltonian and the vacuum oscillation frequencies and two-body couplings are chosen arbitrarily, phases which contribute to the coherence of trace-reduced partitions of the many-body system may average to zero
in the mass basis.  If this 
occurs, then few-body RDMs will be approximately statistically mixed states amenable to description by a few parameter statistical 
distribution. This dephasing in the mass basis accompanied by relaxation to an asymptotic state that can be well described using a 
statistical ensemble is typical of chaotic systems undergoing thermalization~\cite{Srednicki1994,Rigol_2008}. 

In our setup, the 
presence of a global conserved charge $\langle J_z\rangle$ suggests that a natural statistical description of the subsystems would be 
a grand canonical ensemble. In the special case of uniform couplings, the system becomes integrable and the extensive number of conserved 
charges can be accommodated in principle using the Generalized Gibbs Ensemble~\cite{Rigol_2007,Vidmar_2016}.
For the RC case the one-body RDM can be approximated by a Boltzmann distribution of the form
\begin{equation} \label{eq:statFit_1}
    \rho_{i}(\beta,\mu) = \frac{   e^{\beta \omega_{i}/2 + \mu} \qmouter{\nu_{1}}{\nu_{1}} +
        e^{-\beta \omega_{i}/2 - \mu} \qmouter{\nu_{2}}{\nu_{2}} }{2 \cosh(\beta \omega_{i}/2 + \mu)},  
\end{equation}
where the chemical potential $\mu$ and temperature $\beta^{-1}$ are constrained by fixing the total average $\langle \sigma_{z} \rangle$ and the 
average energy $\langle H_{\text{vac}} \rangle$ at late times. The $\nu_{1}$ probability for the statistical mixture state is shown 
with a dashed green line in both panels of Fig.~\ref{fig:obFlavorOps}. For the integrable model (UC) the presence of an extensive set of conserved 
charges prevents dephasing in the mass basis and the predictions obtained from Eq.~\eqref{eq:statFit_1} deviate significantly from the 
correct (MB) results. On the other hand, for the RC model the MB results are successfully captured by the statistical ansatz. 

\begin{figure}
    \centering
    \includegraphics[scale=0.34]{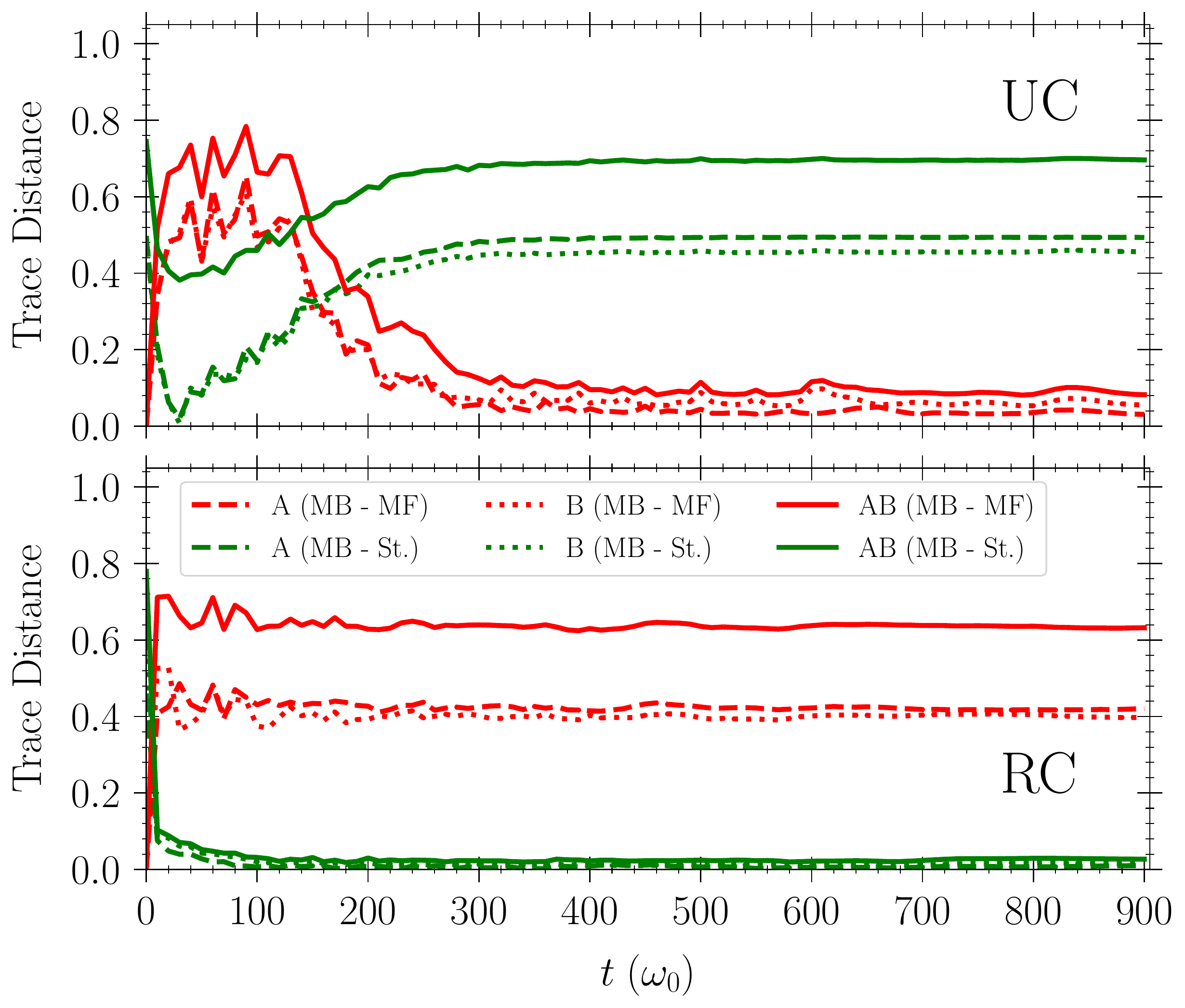}
    \caption{
    Trace distances between the one-body RDMs in the MB and (time-dependent) MF calculations and the (late-time fit) statistical mixed state (St.) for two individual neutrinos $A$ and $B$ (marked with the gold asterisks 
    in Fig.~\ref{fig:obFlavorOps} for which $\omega_{A} < \omega_{B}$), respectively. Also plotted are the trace distances between the corresponding two-body RDMs with both $A$ and $B$. 
    }
    \label{fig:trDistances}
\end{figure}

To quantify the differences between the statistical mixture state, the MF state, and the true evolution of the MB state, we show 
in Fig.~\ref{fig:trDistances} the trace distance 
$T(\rho,\rho')=\frac{1}{2}\text{Tr}\left( \sqrt{ (\rho-\rho')^{\dagger} (\rho-\rho') } \right)$ between the RDMs for two 
representative neutrinos at both the one- and two-body level.  The trace distance $T(\rho,\rho')$ characterizes the maximum observed 
deviation for general observables measured in either state. As can be seen from these results, at late times for the RC system the 
one and two-body RDMs can be approximated with a good accuracy by fitting Eq.~\eqref{eq:statFit_1} to the late time average energy and $J_{z}$ polarization of the true MB state. The same does not hold in the UC 
case possibly due to integrability preventing dephasing and relaxation to such state. The red curves show the distance between the MF and MB one- and two-body RDMs at all times, and they show substantial differences particularly at early times for both choices of two-body couplings. 

{\it Conclusion---}
In this work, we have provided evidence that a general (not fully polarized) initial product state under the action of 
the considered time-dependent, all-to-all coupled interaction Hamiltonian with non-uniform one- and two-body couplings quickly 
develops entanglement among many-particle subsystems.  This entanglement, coupled with the dense avoided level crossings in the 
time dependent Hamiltonian, leads to a strong dephasing in the mass basis and produces final states whose few-neutrino reduced 
density matrices are well described by statistical distributions. This relaxation dynamically leads to a loss of coherence between 
the mass states of individual neutrinos with the net effect of substantially reducing the amplitude of further vacuum flavor oscillations.

These results point to several important aspects of neutrino flavor transport models that warrant further attention including the 
exploration of the role played by coupling neutrinos with external matter (a possibly important term neglected in this first study), 
the possibility of improving the MF prediction using semi-classical approaches (as e.g. the one proposed in Ref.~\cite{Lacroix:2022krq}) 
and the connection between the efficient dephasing observed here to dynamical phase transitions observed in static models~\cite{Roggero2021b,Roggero2022}.

\acknowledgments{
We thank Duff Neill for productive conversation regarding quantum chaotic systems.
This work was supported by the Quantum Science Center (QSC), a National Quantum Information Science Research Center of the U.S. Department of Energy (DOE) and by the U.S. Department of Energy, Office of Science, Office of Nuclear Physics (NP). H.~D.\ is supported by the US DOE NP grant No.\ DE-SC0017803 at UNM. 
}

\bibliography{refs.bib}

\end{document}